\documentclass[conference]{IEEEtran}

\usepackage[utf8x]{inputenc}
\usepackage[OT1]{fontenc}
\PassOptionsToPackage{unicode}{hyperref}

\usepackage[romanian,english]{babel}

\usepackage{graphicx}
\graphicspath{{img/}}

\usepackage{listings}
\usepackage{multirow}
\usepackage{longtable}

\usepackage{url}
\usepackage{color}

\usepackage{topcapt}

\title{Design and Implementation of a High Quality and High Throughput TRNG in FPGA}
\author{%
Cristian KLEIN$^1$\\
Technical University of Cluj-Napoca\\
E-mail: cristi@net.utcluj.ro
\and
Octavian CRET$^2$\\
Technical University of Cluj-Napoca\\
E-mail: Octavian.Cret@cs.utcluj.ro
\and
Alin SUCIU$^2$\\
Technical University of Cluj-Napoca\\
E-mail: Alin.Suciu@cs.utcluj.ro
}

\begin{document}

\maketitle
\footnotetext[1]{Financiar support from INRIA is hereby greatly acknowledged.}
\footnotetext[2]{This work was supported by the CNMP funded CryptoRand project, nr. 11-020/2007.}

\begin{abstract}
This paper focuses on the design and implementation of a high-quality and high-throughput true-random number generator (TRNG) in FPGA. Various practical issues which we encountered are highlighted and the influence of the various parameters on the functioning of the TRNG are discussed. We also propose a few values for the parameters which use the minimum amount of the resources but still pass common random number generator test batteries such as DieHard and TestU01.
\end{abstract}

\section{Introduction}
Random numbers are at the very core of cryptographic algorithms. They are used to generate either the public / private key pair in asymmetric algorithms, or the shared secret / initialisation vector in symmetric cyphers. The ability of an adversary to predict the random numbers used, voids 
the security of the cypher. In fact, the only cypher whose security is proven to be perfect (one time pad) relies on the fact that the random number source is perfectly uniform and unpredictable.

Random number generators are of two types. The first one, \emph{pseudo-random number generators} (PRNG), are the ones in which a person who designed the system, or has access to its internal state can predict the next random number. The system is a deterministic Finite State Machine, whose evolution can usually be described based on an arithmetic formula which determines its transition from a given internal state to another state, while outputting a random number based on a portion of the state. They have the advantage of having high speeds and some of them are cryptographically secure. However, all of them require an initial state (also called \emph{seed}), which determines the sequence of numbers which will be generated. The importance of well seeding a pseudo-random number generator has recently been highlighted in a Debian security vulnerability\cite{De08}.

The second type of random numbers generators are \emph{true-random number generators} (TRNG), whose output cannot be predicted, not even by the person who designed them. They are usually based on sampling some kind of physical phenomenon (such as noise) which has a lot of randomness. Although one would be tempted to use only TRNGs in cryptography, their smaller throughput prohibits this, so they are commonly used to seed PRNGs.

FPGAs are becoming a popular choice for implementing cryptographic devices, due to the fact they represent the middle ground between the flexibility of the microprocessor and the speed of an ASIC. They allow creating high-throughput cryptographic devices while at the same time making it possible to change or improve the underlying algorithms, should a security flaw be discovered.

Many papers \cite{Ko04}\cite{Su06}\cite{Sc06}\cite{Ts03}\cite{Si05} have explored the possibility of implementing TRNGs in FPGAs, motivated by the avoidance of additional hardware, and the impossibility to intercept the data stream between the TRNG and the actual cryptographic implementation. While all of them claim to obtain good-quality TRNG, few mention explicitly the methods involved in transforming a hardware which is supposed to work predictably into a source of entropy.

This paper elaborates on the design and implementation of the TRNG principle presented by Martin and Stinsonin in \cite{Su06} and highlights a few practical issues encountered while implementing a high-quality TRNG based on it. We identified a few generic parameters, whose influence on the TRNG will also be presented in this paper. The ultimate purpose is to enable the reader to easily implement this TRNG on a low-cost FPGA development board, such as one featuring a Xilinx Spartan 3E.

\section{Principle}

\begin{figure}
\centering
\includegraphics[height=5cm]{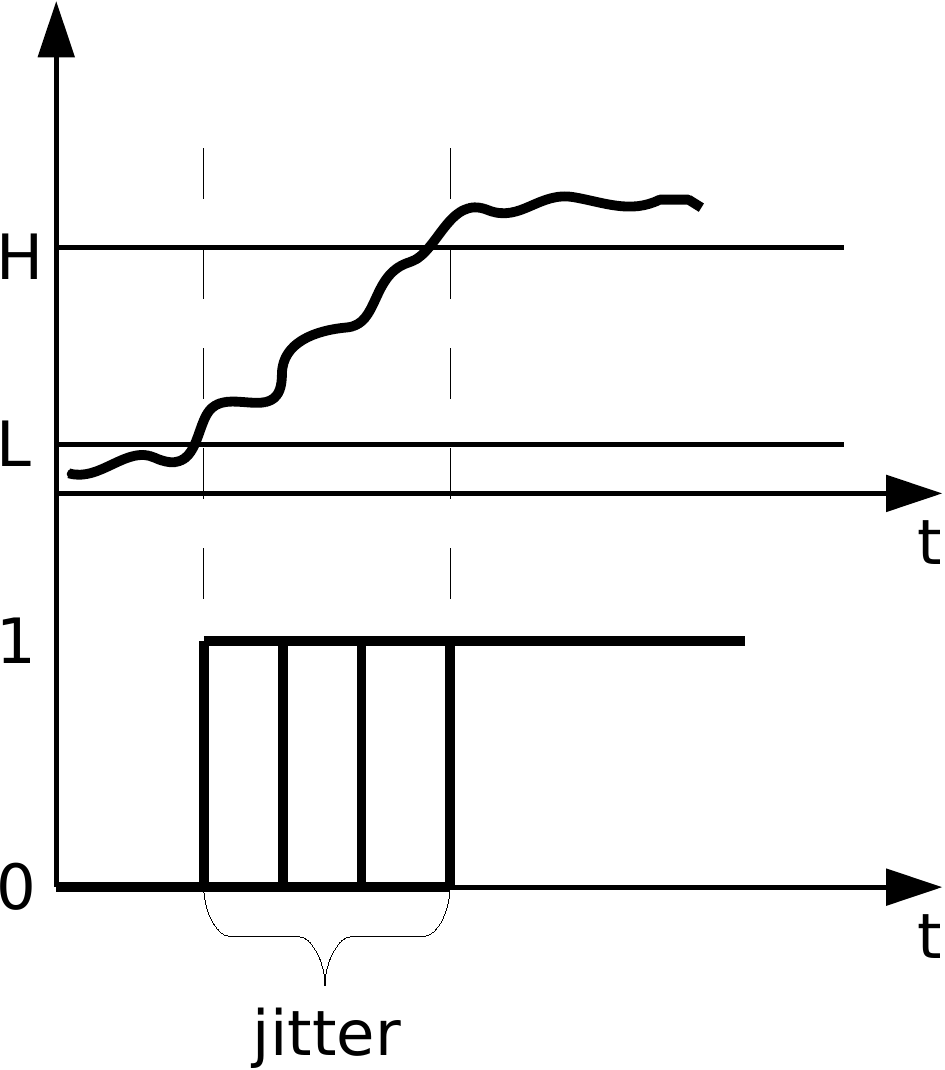}
\caption{The Jitter Model}
\label{fig:jitter}
\end{figure}

Like many TRNG implemented in FPGA, this design is based on sampling jitter. In essence, due to various noise sources such as that induced by the power supply but also by nearby components, the behaviour of "demanding a 0 or 1" from the transition slope of an output is unpredictable. This is caused by the fact that each technology defines a low (L) threshold, which is the upper limit for voltages which represent a logic 0, and a high (H) threshold which is the lower limit of logic 1. Output behaviour between these two values is not well defined. This can be modelled as if the output of the component would have a perfectly vertical slope, but the time of the transition is unknown and can range from the beginning until the end of the real slope (figure~\ref{fig:jitter}).

\begin{figure}
\centering
\includegraphics[width=5cm]{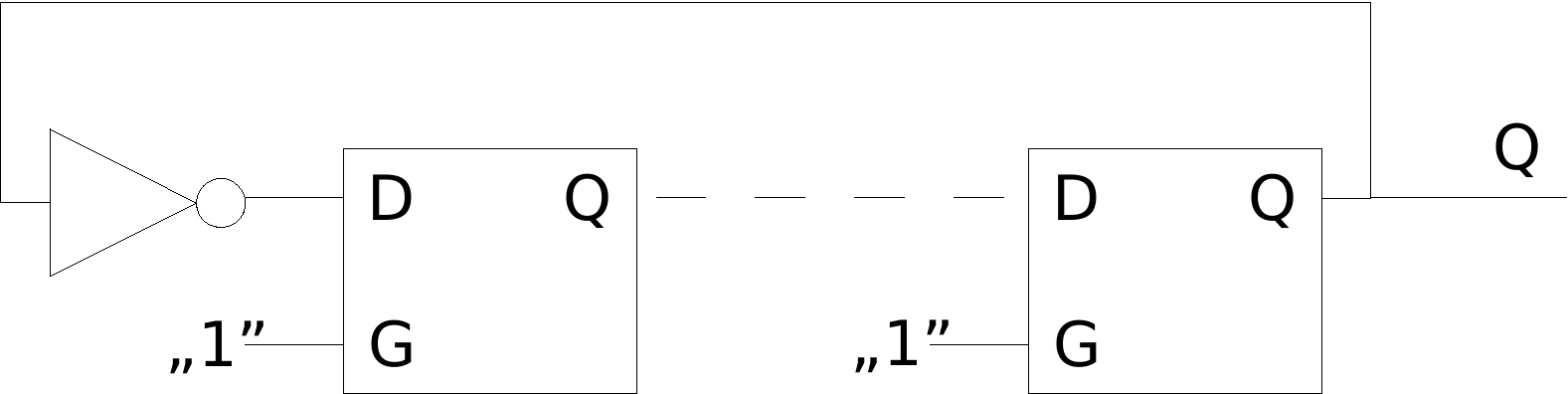}
\caption{Ring Oscillator (RO)}
\label{fig:ro}
\end{figure}

In order to produce jitter, TRNGs employ one or more ring oscillators (RO) (figure~\ref{fig:ro}). These are composed of a ring of odd number of inverting elements and an arbitrary number of delaying elements. The simplest RO is composed of a single inverter and a buffer. The output of a RO is never stable and does transitions from 0 to 1 and back to 0 with a frequency given by the propagation delay of the constituting elements. Due to the above described phenomena, the period of an oscillation will not be constant, because it will vary by a small amount each time. This is the manifestation of jitter and the source of entropy which our TRNG will collect.

Our first attempt was to create a TRNG based on \cite{Ko04}, which uses one RO to sample the output of the other RO. We appreciated this approach due to the fact that the whole stream before the post-processing phase is random (although it might be biased a little bit). We also favoured this design, because, if there is some kind of predictable jitter (such as coming from the power source), both ROs are influenced the same way which should cancel out at the sampler. However, we found that putting this design into practice is very challenging. The two ROs have to be nearly identical, which requires manual placing and routing. Even after achieving that, the design proved to be very sensitive to other components in the FPGA. At the time of this writing we have been able to create a TRNG which outputs very good numbers on a serial interface, but have been unable to obtain good quality random numbers at the TRNG's highest speed.

Therefore, we chose to implement \cite{Su06} which uses multiple ROs whose outputs are XOR-ed. A flip-flop whose clock is driven by a fixed frequency will sample the combined output of the ROs. The obtained stream will hit both jitter zones (our source of entropy) and flat zones (which are highly predictable). A post-processing phase is required which consists in a resilience function\cite{Resilience}. In essence, the function takes an $m$-bit input, out of which $n$-bits are known to be random (but we can't determine which ones) and outputs $n$-bits which are known to be random. For $n=1$, the simplest resilience function is to xor all the input bits. Suppose all but one bits are deterministic, but the probability of a 0 or 1 value of one bit are equal, the output of the xor will also have equal probability of being 0 or 1.

\section{Implementation Issues}

\begin{figure*}[t]
\centering
\includegraphics[width=\textwidth]{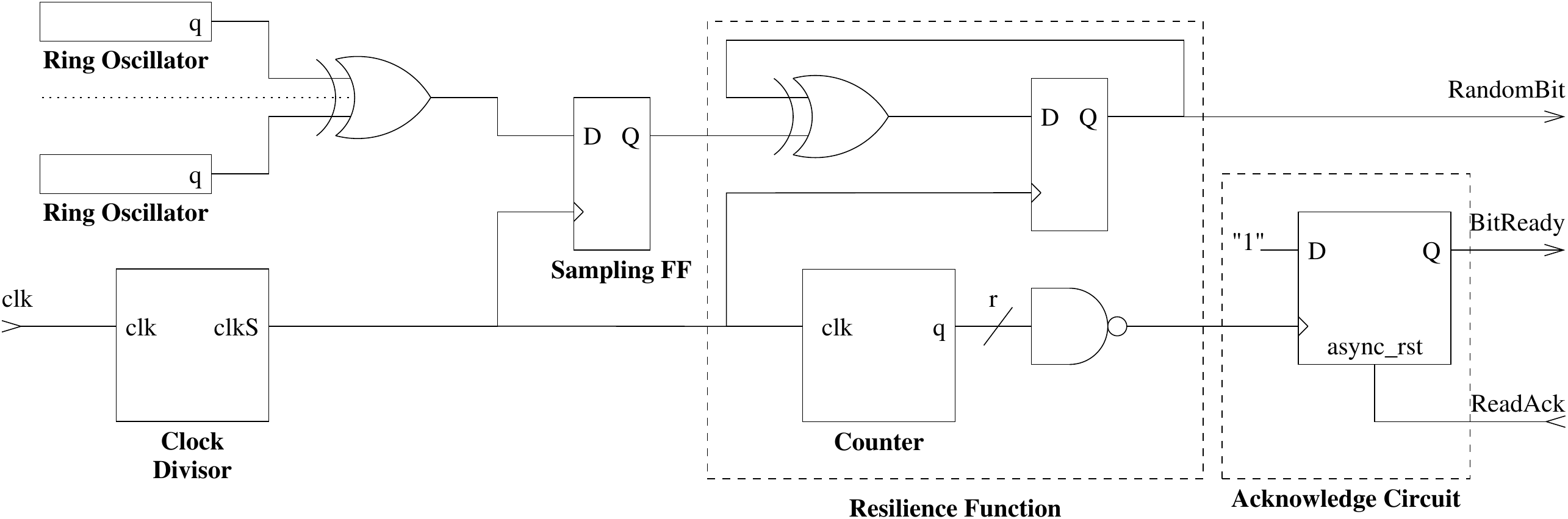}
\caption{TRNG Scheme}
\label{fig:trng}
\end{figure*}

\subsection{Creating ROs in VHDL}
Our first goal was to create a VHDL component which would implement a ring oscillator with a parametrised length.

We first studied what resources are available in the FPGA to create ring oscillators. The main building block of the FPGA, the CLB are the only ones that actually contain logic, and are interconnected by a network of routing wires. The CLB contains a LUT, an invertor and a memory element which can be either used as a latch or as a flip-flop. The output of latch / FF goes directly out of the CLB into the interconnection network. Two CLBs are grouped together in a slice, however in order to connect the output of one latch / FF to the other CLB in the same slice, the wire has to exit the slice, go through the interconnection network and reenter the CLB. From Xilinx's reports we noticed that the main delays in FPGA come from latches and routing. The inverter induces a negligible delay. Another interesting thing we noticed is that during the mapping phase, a {\tt GLOBAL\_LOGIC1} signal is created which provides logic "1" for all the CLBs that require it.

Having the knowledge above, we chose to have a single inverter at the beginning of the chain and a variable number of latches as delay components (like in figure~\ref{fig:ro}). A single inverter allows us to create ROs which both even and odd number of latches. By default Xilinx's synthesis tool optimises out all but one latch, due to the fact that they seem redundant from its viewpoint. In order to prevent this, we must set the "keep" attribute\cite{XiKeep} of the {\tt d} bus which interconnects the latches:

\begin{verbatim}
attribute keep : string;
attribute keep of d : signal is "true";
\end{verbatim}

This tells the synthesis and mapping tool that we want the individual {\tt d} signals not to be absorbed into a CLB. Each of them must pass through the interconnection network, which forces the tools to map the redundant latches to CLBs.

To make sure that the inverter does not add more delay, we added the {\tt not} keyword directly into the port map of the first latch, without assigning it a signal. This has the effect that the inverter and the first latch are mapped to the same CLB.

\subsection{Sampler}
We chose to give the whole TRNG circuit the same interface as the one used by \cite{Ko04}, to which we added an input clock signal (figure~\ref{fig:trng}). The {\tt BitReady} output signal is high when the TRNG has a new random bit, which will appear at the {\tt RandomBit} pin. When the external circuit has stored the random bit, it will acknowledge the TRNG by rising the {\tt ReadAck} pin.

Although our particular TRNG is synchronous, all three signals are assumed to be asynchronous, both inside the TRNG and the external circuitry that connects to it. We took this decision for two purposes: first, we wanted to be able to use a RO's output as the sampling clock, which would make the TRNG truly asynchronous, and secondly, we wanted to used the very same design to test future TRNG, which might be asynchronous in nature.

\subsection{Resilience Function}
Contrary to the design employed by others, we chose as the resilience function a simple XOR of $2^r$-bits (where $r$ is a generic parameter). We did this because we feared that using a more complex resilience function may hide possible defects in our TRNG, which we obviously want to avoid. Moreover, some resilience functions (such as cyclic codes) are implemented using shift registers and XORs which might act as PRNG. We specifically want to test how well the TRNG works with minimal post-processing. Using the TRNG to seed a PRNG (although a possibly weak one) is against the purpose of our paper.

\subsection{High-throughput Measurements}

It was very important for us to validate the TRNG at its maximum speed. We feared that the output interface from the FPGA to the computer (where the random bits are collected and analysed), whether RS232 or USB, would do additional sampling of the (possible partially) random stream. This would return more optimistic results compared to the TRNG being used only inside the FPGA.

\begin{figure*}
\centering
\includegraphics[width=0.8\textwidth]{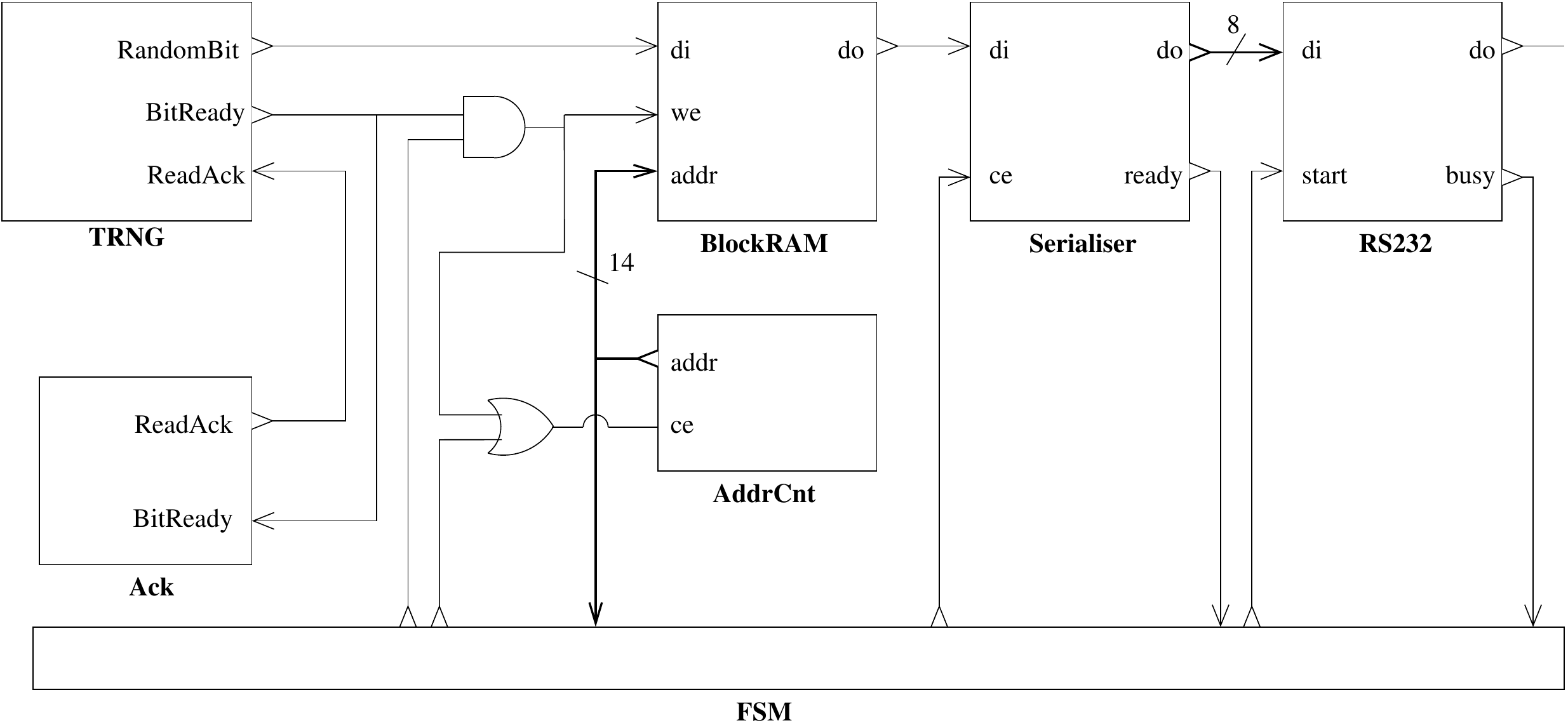}
\caption{High-throughput Measurement Scheme}
\label{fig:hbm}
\end{figure*}

In order to achieve this, we created a design which would first fill a 16 Kbit BlockRAM with TRNG output, then transfer this to the output interface (figure~\ref{fig:hbm}). We think that this is very close to how a TRNG would be used in a FPGA cryptographic application: the cypher gets values from the entropy buffer and while the algorithm proceeds, the TRNG fills back the entropy buffer. 

The design is able to handle burst transfers from the TRNG. The data-in port of the RAM is directly connected to the TRNGs output. The control signals of the address counter and the write-enable port of the RAM are directly connected to the {\tt BitReady} port of the TRNG, provided the FSM is in the {\tt FillRAM} state. A separate circuit is used to drive the {\tt ReadAck} port of the TRNG which sets it to $1$ at the very next clock rising edge, exactly when the RAM has stored the random bit.

\begin{figure*}[!t]
\centering
\includegraphics[width=0.8\textwidth]{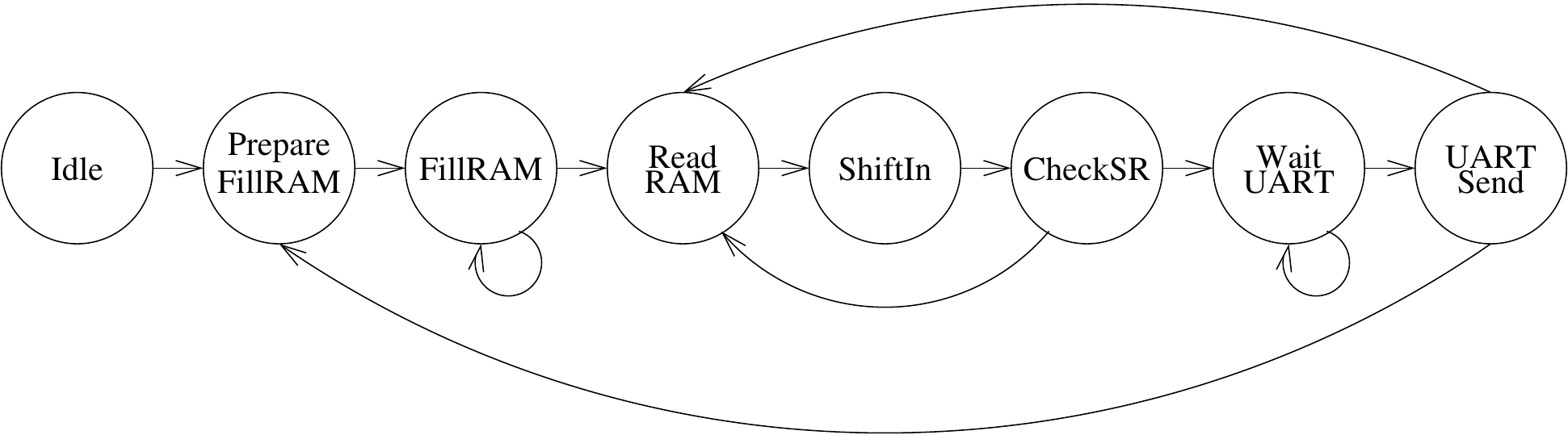}
\caption{FSM State Diagram}
\label{fig:fsm}
\end{figure*}

The FSM which controls this circuit has 8 states (figure~\ref{fig:fsm}). The first, {\tt Idle} is the state in which the FSM is set immediately after reset. Transition is made immediately to the {\tt PrepareFillRAM} state, which resets the address counter. Next, the {\tt FillRAM} state allows the counter to increase and the RAM to store values when a new random bit is ready. The FSM stays in this state until the RAM is filled (i.e. the RAM address counter wraps around). The next three states ({\tt ReadRAM}, {\tt ShiftIn}, {\tt CheckSR}) serialise the bits stored in the RAM into a byte for being transmitted to the UART module. The same counter is used to control the address of the RAM, but it is only incremented in the {\tt ShiftIn} state. Finally, when a byte is complete (i.e. the {\tt Serialiser} sets the {\tt ready} port to 1) the FSM will wait for the UART to complete the previous transmission ({\tt WaitUART}), then dispatch the data ({\tt UARTSend}). If there is more data to transmit (i.e. RAM address counter is non-zero) then the FSM will transition to the {\tt ReadRAM} state, serialising the next byte. If the whole contents of the RAM has been transmitted, it will be freshly filled with random numbers, by jumping to the {\tt PrepareFillRAM} state.

\section{Tuning the TRNG}
A very important practical aspect of the TRNG is to know the influence of its generic parameters on the quality of its output. We also wanted to test practically what is the smallest number of FPGA resources which are required for this TRNG. We used the DieHard\cite{DieHard} and the TestU01\cite{TestU01} (NIST, Rabbit and Alphabit battery) suites to test the quality of the TRNG output. We only considered parameters for which the output of the TRNG passed all tests, i.e. all DieHard p-values are different from 0 or 1, and TestU01 prints "All tests were passed". All files which we downloaded had at least 10 MB, due to limitations in DieHard. Interestingly, the TestU01 library proved to be a lot more sensitive than DieHard.

The proposed TRNG has the following generic parameters: number of ring oscillators ($n$), length of ring oscillators ($l$), sampling frequency divisor ($2^d$) and resilience function input width ($2^r$).

The first two aspects in which we were interested is the throughput and the amount of resources this design uses. The throughput can be easily computed as the output rate of the TRNG is the input clock frequency, divided first by the clock divider, then the resilience function. The formula is:

\begin{equation}
b = \frac{f}{2^r * 2^d}
\end{equation}
where $f$ is the input clock frequency of the TRNG and $b$ is the throughput in bps.

The amount of resources can also be easily estimated. Each RO uses $l$ CLBs. The xor stage is synthesised as a tree of LUTs. Due to the fact that the slice of a Spartan 3E device contains 4-bit input LUTs the number of CLBs is $ \left\lceil \frac{n-1}{3} \right\rceil $. The clock divisor uses approximately $\frac{d}{4}$ CLBs. The counter uses about $\frac{r}{4}$ CLBs, while the and stage at its output uses $ \left\lceil \frac{r-1}{3} \right\rceil $. The other components (sampler FF, resilience XOR and FF, acknowledge circuit) use 3 CLBs. Therefore the total number of used CLBs ($C$) is:

\begin{equation}
C = l + \left\lceil \frac{n-1}{3} \right\rceil + \frac{d}{4} + \frac{r}{4} + \left\lceil \frac{r-1}{3} \right\rceil + 3
\end{equation}

During our experiments we concluded that the quality of the output random bit stream increases with the increase of $d$, $r$ and $n$. As the number of ring oscillators ($n$) increases and because the ring oscillators don't have the exactly same frequency, the signal after xoring them will be composed of much more jitter than flat zones. This means that the sampler will return much more non-deterministic bits compared to the amount of deterministic bits. The more input bits the resilience function has the more non-deterministic bits will be xored with the deterministic bits, which in effect will increase the chance of the TRNG to output a truly random bit.

%{\color{magenta} De ce cre?te odat? cu $d$? Eu am zis a?a:}
Regarding the clock divider, if $d$ is too small (even comparable to the frequency of the ring oscillators), the sampler tends to hit the same flat zone or return the same non-deterministic bit several times. The resulting correlated bits can of course be eliminated in the resilience stage, provided that $r$ is large enough. We can clearly see that the well-known throughput vs. resources conflict also holds in case of this TRNG.

We haven't found any significant influence of $l$ on the quality of the random numbers. This might be due to the fact that while each delay element increases the output period of the ring oscillators, it also increases the amount of jitter, so the percentage of the jitter after the sampling stage remains roughly the same. Although one is tempted to use ring oscillators with the minimum length, we recommend to use $l\ge3$ to make sure that the system does not remain without jitter in extreme conditions such as sudden temperature variations.

In our experiments the parameters values presented in table~\ref{tab:param} created a TRNG which passed all tests, while minimizing the number of ring oscillators. Please note that in case one wants to be absolutely sure that the TRNG will output high-quality random numbers, higher values should be used for $r$ or, if bandwidth is an issue, $n$.

\begin{table}
\centering
\topcaption{Parameters for High Quality TRNG}
\label{tab:param}
\begin{tabular}{cccc|r}
$d$ & $r$ & $n$ & $l$ & \parbox{2cm}{\centering throughput \newline (Kbps)} \\
\hline
0 & 2 & 20 & 3 & 12500 \\
0 & 3 & 10 & 3 & 6250 \\
2 & 2 & 10 & 3 & 3125 \\
5 & 3 & 5 & 3 & 195 \\
\hline
\end{tabular}
\end{table}

\section{Speeding up the TRNG}
FPGAs are becoming large enough to allow massive pipelining of arithmetic operands and compute one result per clock. In some applications it might be desirable to generate random numbers at the maximum frequency of the FPGA. In the above design, both the resilience function (characterised by $r$) and the sample clock divider ($d$) lower the frequency of the TRNG. While we could set $d$ to zero, so that the sampling clock is set to maximum, we can never set $r$ to zero, while at the same time obtain good quality random numbers.

First solution which would come to one's mind is to use multiple parallel TRNGs and multiplex their outputs. Suppose the sample clock divider is equal to zero, each TRNG would output one bit each $2^r$ cycles. This means that we would need $2^r$ TRNGs for generating one random bit on each FPGA clock. While this solution would surely work (due to the fact that by interleaving truly random streams one obtains another truly random stream), we wanted to find a design that would minimise the resource utilisation.

Our idea is that we require the resilience function because not all our bits are sampled from jitter. The same would apply if we would XOR bits coming from different samplers. This way, we would save $2^r$ counters, FFs and AND gate and replace them with one big XOR.

Indeed, we have practically validated the fact that good quality random numbers are generated using the above concept, for $8$ samplers and $20$ ROs / sampler. Interestingly, the number of samplers required is equal to the number of bits which enters the resilience function in the design presented in figure~\ref{fig:trng}.

\newpage % Manually balance last two columns.
Note however, what for the mentioned values, we used $160$ ROs, eight times more. An interesting question is whether this number of ROs could be used to generate a random bit stream, without using a resilience function ($d=0$ and $r=0$ in figure~\ref{fig:trng}). We have practically shown that this is not possible, as explained in \cite{Sc06}. In essence, the probability of sampling a random bit increases with the number of ROs, but never reaches 1. The small percent of the resulting correlated bits is enough to make the TRNG fail quality tests.

\section{Conclusion}

In this paper we have shown how a simple yet of high-quality and high-throughput TRNG can be implemented on a low-end Xilinx Spartan 3E FPGA and presented the main implementation issues one might encounter. We have also discussed the various parameters of the TRNG and the influence they have on the design. We believe that this paper has paved the way to implementing secure cryptographic applications in low-end FPGAs, without requiring any external component.

%\IEEEtriggeratref{3}
\bibliographystyle{abbrv}
\bibliography{main}

\end{document}